\documentclass{article} % For LaTeX2e
\usepackage{nips10submit_e,times}
\usepackage{amsmath}
\usepackage{color}
\usepackage{url}
\usepackage{xspace}
\usepackage[dvips]{graphicx}
\usepackage{subfigure}
\usepackage{multirow}
\usepackage{arydshln}

\newcommand{\xhdr}[1]{\textbf{#1}.}
\newcommand{\T}{{\tau}}
\newcommand{\Tic}{{\T_i^c}}
\newcommand{\SI}{{\textsc{SI}}\xspace}
\newcommand{\SIS}{{\textsc{SIS}}\xspace}
\newcommand{\SIR}{{\textsc{SIR}}\xspace}

\title{On the Convexity of Latent Social Network Inference}

\author{
Seth A. Myers \\
Institute for Computational  \\and Mathematical Engineering\\
Stanford University\\
\texttt{samyers@stanford.edu} \\
\And
Jure Leskovec \\
Department of Computer Science \\
Stanford University\\
\texttt{jure@cs.stanford.edu} \\
}

\nipsfinalcopy % Uncomment for camera-ready version

\begin{document}

\maketitle

\begin{abstract}
In many real-world scenarios, it is nearly impossible to collect explicit
social network data. In such cases, whole networks must be inferred from
underlying observations. Here, we formulate  the problem of inferring latent
social networks based on network diffusion or disease propagation data. We
consider contagions propagating over the edges of an unobserved social network, where we
only observe the times when nodes became infected, but not who infected them.
Given such node infection times, we then identify the optimal network that best
explains the observed data. We present a maximum likelihood approach based on
convex programming with a $l_1$-like penalty term that encourages sparsity.
Experiments on real and synthetic data reveal that our method near-perfectly
recovers the underlying network structure as well as the parameters of the contagion
propagation model. Moreover, our approach scales well as it can infer optimal
networks of thousands of nodes in a matter of minutes.
\end{abstract}

\section{Introduction}
\label{sec:intro}
Social network analysis has traditionally relied on self-reported data
collected via interviews and questionnaires~\cite{wasserman94book}. As
collecting such data is tedious and expensive, traditional social network
studies typically involved a very limited number of people (usually less than
100).
The emergence of large scale social computing applications has made
massive social network data~\cite{jure08messenger} available, but there
are important settings where network data is hard to obtain and thus the whole
network must thus be inferred from the data. For example, populations, like
drug injection users or men who have sex with men, are ``hidden'' or
``hard-to-reach''. Collecting social networks of such populations is near
impossible, and thus whole networks have to be inferred from the
observational data.

Even though inferring social networks has been attempted in the past, it usually
assumes that the pairwise interaction data is already available ~\cite{eagle09inferring}. In this case, the
problem of network inference reduces to deciding whether to include the
interaction between a pair of nodes as an edge in the underlying network.
For example, inferring networks from pairwise interactions of cell-phone
call~\cite{eagle09inferring} or
email~\cite{dechoudhury10interpersonal,kossinets06empirical} records simply
reduces down to selecting the right threshold $\tau$ such that an edge
$(u,v)$ is included in the network if $u$ and $v$ interacted more than $\tau$
times in the dataset. Similarly, inferring networks of interactions between
proteins in a cell usually reduces to determining the right
threshold~\cite{giot2003protein,middendorf2005inferring}.

We address the problem of inferring the structure of unobserved social networks
in a much more ambitious setting. We consider a diffusion process where a
contagion (e.g., disease, information, product adoption) spreads over the edges of the
network, and all that we observe are the infection times of nodes, but not who
infected whom i.e. we do not observe the edges over which the contagion spread. The goal then is to reconstruct the underlying social network
along the edges of which the contagion diffused.

We think of a diffusion on a network as a process where neighboring nodes switch
states from inactive to active. The network over which activations propagate is
usually {\em unknown} and {\em unobserved}. Commonly, we only observe the times
when particular nodes get ``infected'' but we {\em do not} observe {\em who}
infected them. In case of information propagation, as bloggers discover new
information, they write about it without explicitly citing the
source~\cite{jure09memes}. Thus, we only observe the time when a blog gets
``infected'' but not where it got infected from. Similarly, in disease
spreading, we observe people getting sick without usually knowing who infected
them~\cite{wallinga04epidemic}. And, in a viral marketing setting, we observe
people purchasing products or adopting particular behaviors without explicitly
knowing who was the influencer that caused the adoption or the
purchase~\cite{hill06viral}. Thus, the question is, if we assume that the network is static over time, is it possible to
reconstruct the unobserved social network over which diffusions took place?
What is the structure of such a network?

We develop convex programming based approach for inferring the latent social
networks from diffusion data. We first formulate a generative probabilistic
model of how, on a fixed hypothetical network, contagions spread through the
network. We then write down the likelihood of observed diffusion data under a
given network and diffusion model parameters. Through a series of steps we show
how to obtain a convex program with a $l_1$-like penalty term that encourages
sparsity. We evaluate our approach on
synthetic as well as real-world email and viral marketing datasets. Experiments
reveal that we can near-perfectly recover the underlying network structure as
well as the parameters of the propagation model. Moreover, our approach scales
well since we can infer optimal networks of a thousand nodes in a matter of
minutes.

\xhdr{Further related work} There are several different lines of work connected
to our research. First is the network structure learning for estimating
the dependency structure of directed graphical models~\cite{getoor2003learning}
and probabilistic relational models~\cite{getoor2003learning}. However, these
formulations are often intractable and one has to reside to heuristic
solutions. Recently, graphical Lasso
methods~\cite{wainwright06graphical,schmidt2007learning,friedman08lasso,meinshausen2006high}
for static sparse graph estimation and extensions to time evolving graphical
models~\cite{ahmed2009recovering,ghahramani1998learning,song09timevarying} have
been proposed with lots of success. Our work here is similar in a sense that we
``regress'' the infection times of a target node on infection times of other
nodes. Additionally, our work is also related to a link prediction
problem~\cite{janse03linkpred,taskar03linkpred,libennowell03linkpred,vert2005supervised}
but different in a sense that this line of work assumes that part of the
network is already visible to us.

The work most closely related to ours, however, is ~\cite{gomez10netinf}, which also infers networks through cascade data.  The algorithm proposed (called NetInf) assumes that the weights of the edges in latent network are homogeneous, i.e. all connected nodes in the network infect/influence their neighbors with the same probability.  When this assumption holds, the algorithm is very accurate and is computationally feasible, but here we remove this assumption in order to address a more general problem.  Furthermore, where ~\cite{gomez10netinf} is an approximation algorithm, our approach guarantees {\em optimality} while easily
handling networks with thousands of nodes.

%\section{Related Work}
%\label{sec:related}
%\input{020related}

\section{Problem Formulation and the Proposed Method}
\label{sec:proposed}
We now define the problem of inferring a latent social networks based on
network diffusion data, where we only observe identities of infected nodes.
Thus, for each node we know the interval during which the node was infected,
whereas the source of each node's infection is unknown.  We assume only that an infected node was previously infected by some other previously infected node to which it is connected in the latent social network (which we are trying to infer).
Our methodology can handle a wide class of information diffusion and epidemic
models, like the independent contagion model, the Susceptible--Infected (\SI),
Susceptible--Infected--Susceptible (\SIS) or even the
Susceptible--Infected--Recovered (\SIR) model~\cite{bailey75mathematical}.
We show that calculating the maximum likelihood estimator (MLE) of the latent
network (under any of the above diffusion models) is equivalent to a convex
problem that can be efficiently solved.

\xhdr{Problem formulation: The cascade model}
We start by first introducing the model of the diffusion process. As the
contagion spreads through the network, it leaves a trace that we call a {\em
cascade}.
Assume a population of $N$ nodes, and let $A$ be the $N\times N$ weighted
adjacency matrix of the network that is unobserved and that we aim to infer.
Each entry $(i,j)$ of $A$ models the conditional probability of infection
transmission:
\begin{align*}
  A_{ij}=P(\mbox{node $i$ infects node $j$} \ |  \ \mbox{node $i$ is infected}).
\end{align*}
The temporal properties of most types of cascades, especially disease spread, are governed by a transmission (or incubation) period.  The transmission time model $w(t)$ specifies how long it takes for
the infection to transmit from one node to another, and the recovery model
$r(t)$ models the time of how long a node is infected before it recovers.
Thus, whenever some node $i$, which was infected at time $\T_i$, infects another
node $j$, the time separating two infection times is sampled from $w(t)$, i.e.,
infection time of node $j$ is $\T_j=\T_i+t$, where $t$ is distributed by
$w(t)$. Similarly, the duration of each node's infection is sampled from
$r(t)$. Both $w(t)$ and $r(t)$ are general probability distributions with
strictly nonnegative support.

A cascade $c$ is initiated by randomly selecting a node to become infected at
time $t=0$. Let $\T_i$ denote the time of infection of node $i$.  When node $i$ becomes infected, it infects each of its neighbors independently in the network, with probabilities governed by $A$.  Specifically, if $i$ becomes infected and $j$ is susceptible, then $j$ will become infected with probability $A_{ij}$.  Once it has been determined which of $i$'s neighbors will be infected, the infection time of each newly infected neighbor will be the sum of $\tau_i$ and an interval of time sampled from $w(t)$.  The transmission time for each new infection is sampled independently from $w(t)$.

Once a node becomes infected, depending on the model, different scenarios happen. In the \SIS model,
node $i$ will become susceptible to infection again at time $\T_i+r_i$. On the
other hand, under the \SIR model, node $i$ will recover and can never be
infected again. Our work here mainly considers the \SI model, where
nodes remain infected forever, i.e., it will never recover, $r_i=\infty$.  It is important to note, however, that our approach can handle all of these models with almost no modification to the algorithm.

For each cascade $c$, we then observe the node infection times $\Tic$ as well
as the duration of infection, but the source of each node's infection remains
hidden.  The goal then is to, based on observed set of cascade infection times
$D$, infer the weighted adjacency matrix $A$, where $A_{ij}$ models the edge
transmission probability.

\xhdr{Maximum Likelihood Formulation} Let $D$ be the set of observed cascades.
For each cascade $c$, let $\Tic$ be the time of infection for node $i$. Note
that if node $i$ did not get infected in cascade $c$, then $\Tic=\infty$. Also,
let $X_c(t)$ denote the set of all nodes that are in an infected state at time $t$
in cascade $c$. We know the infection of each node was the result of an unknown, previously infected node to which it is connected, so the component of the likelihood function for each infection will be dependent on all previously infected nodes.  Specifically, the likelihood function for a fixed given $A$ is
\begin{align*}
  L(A;D) &= \prod_{c\in D} \left[ \left( \prod_{i; \tau_i^c < \infty} P(\mbox{$i$ infected at $\tau_i^c$}| \mbox{$X_c(\tau_i^c)$}) \right) \cdot  \left( \prod_{i;\tau_i^c=\infty}P(i \mbox{ never infected}|X_c(t) \, \forall \, t) \right) \right] \\
  &=\prod_{c\in D} \left[ \left( \prod_{i; \tau_i^c < \infty} \left(1-\prod_{j;\tau_j \leq \tau_i}(1 - w(\tau_i^c -\tau_j^c)A_{ji})\right) \right) \cdot  \left( \prod_{i;\tau_i^c=\infty}\prod_{j;\tau^c_j < \infty}(1 - A_{ji})\right) \right].
\end{align*}
The likelihood function is composed of two terms. Consider some cascade $c$.  First, for every node $i$ that got infected at time $\Tic$ we compute the probability that at least one other previously
infected node could have infected it. For every non-infected node, we
compute probability that no other node ever infected it. Note that we assume that
both the cascades and infections are conditionally independent. Moreover, in
the case of the \SIS model each node can be infected multiple times during a
single cascade, so there will be multiple observed values for each $\Tic$ and
the likelihood function would have to include each infection time in the product sum.  We omit this
detail for the sake of clarity.

Then the maximum likelihood estimate of $A$ is a solution to
$\min_{A}\, -\log(L(A;D))
%  \mbox{subject } & \mbox{ to} \\
%  0 \leq A_{ij} &\leq 1 \, \forall \, i,j.
$
subject to the constraints $0 \leq A_{ij} \leq 1$ for each $i,j$.

Since a node cannot infect itself, the diagonal of $A$ is strictly zero,
leaving the optimization problem with $N(N-1)$ variables.  This makes scaling
to large networks problematic. We can, however, break this problem into $N$
independent subproblems, each with only $N-1$ variables by observing that the
incoming edges to a node can be inferred independently of the incoming edges of
any other node.  Note that there is no restriction on the structure of $A$ (for example, it is not in general a stochastic matrix), so the columns of $A$ can be inferred independently.

Let node $i$ be the current node of interest for which we would like to infer
its incoming connections. Then the MLE of the $i^{th}$ column of $A$
(designated $A_{:,i}$) that models the strength of $i$'s incoming edges, is the
solution to $\min_{A_{:,i}} -\log(L_i(A_{:,i};D))$,
%  \mbox{subject to}& \\
%  0 \leq A_{ji} \leq 1& \, \forall \,j
%
subject to the constraints $0 \leq A_{ji} \leq 1$ for each $j$, and where
\begin{align*}
  L_i(A_{:,i};D) = \prod_{c\in D; \tau_i^c < \infty} \left[1 - \prod_{j;\tau_j \leq \tau_i} \left( 1 - w(\tau_i^c - \tau_j^c) A_{ji} \right) \right]&\cdot \prod_{c\in D; \tau_i^c = \infty}\left[\prod_{j \in c; \tau_j^c < \infty} \left(1 - A_{ji} \right) \right].
\end{align*}
Lastly, the number of variables can further be reduced by observing that if
node $j$ is never infected in the same cascade as node $i$, then the MLE of
$A_{ji}=0$, and $A_{ji}$ can thus be excluded from the set of variables. This
dramatically reduces the number of variables as in practice the true $A$ does
not induce large cascades, causing the cascades to be sparse in the number of
nodes they infect~\cite{jure07viral,jure06influence}.

\xhdr{Towards the convex problem} The Hessian of the log-likelihood/likelihood
functions are indefinite in general, and this could make finding the globally
optimal MLE for $A$ difficult.  Here, we derive a convex optimization problem
that is equivalent to the above MLE problem.  This not only guarantees
convergence to a globally optimal solution, but it also allows for the use of
highly optimized convex programming methods.

We begin with the problem
$  \max_{A_{:,i}}\, L_i(A;D) $
%  \mbox{subject to}& \\
%  0 \leq A_{ji} \leq 1& \, \forall \,j.
subject to $0 \leq A_{ji} \leq 1$ for each $j$. If we then make the change of
variables $B_{ji} = 1 - A_{ji}$ and
$
  \gamma_c=1 - \prod_{j\in X_c(\tau_i^c)} \left( 1 - w(\tau_i^c - \tau_j^c)A_{ji} \right),
$
the problem then becomes
\begin{align*}
  \max_{\gamma_c,B(:,i)}\prod_{c\in D; \tau_i^c < \infty} \gamma_c &\cdot \prod_{c\in D; \tau_i^c = \infty}\prod_{j \in c; \tau_j^c < \infty} B_{ji} \\
  \mbox{subject to}& \\
  0 \leq B_{ji} \leq 1& \, \ \ \forall \,j \\
  0 \leq \gamma_c \leq 1& \, \ \ \forall \,c \\
  \gamma_c + \prod_{j\in X_c(\tau_i^c)} \left( 1 - w_j^c+ w_j^cB_{ji} \right)& \leq 1 \, \ \ \forall \,c. \label{ineq_gamma}
\end{align*}
where we use shorthand notation $w_j^c \equiv w(\tau_i^c - \tau_j^c)$ (note
that $i$ is fixed). Also, note that the last constraint on $\gamma_c$ is an
inequality instead of an equality constraint. The objective function will
strictly increase when either increasing $\gamma_c$ or $B_{ji}$, so this
inequality will always be a binding constraint at the solution, i.e., the
equality will always hold.  The reason we use the inequality is that this turns
the constraint into an upper bound on a posynomial (assuming $w(t) \leq 1 \,
\forall t$).  Furthermore, with this change of variables the objective function
is a monomial, and our problem satisfies all the requirements for a geometric
program.  Now in order to convexify the geometric program, we apply the change of
variables $\hat{\gamma}=\log(\gamma)$ and $\hat{B_{ji}}=\log(B_{ji})$, and take
the reciprocal of the objective function to turn it into a minimization
problem. Finally, we take the logarithm of the objective function as well as the constraints, and we are
left with the following convex optimization problem
\begin{align*}
  \min_{\hat{\gamma}_c,\hat{B}(:,i)} \sum_{c\in D; \tau_i^c < \infty}-\hat{\gamma}_c-&\sum_{c\in D; \tau_i^c = \infty}\sum_{j \in c; \tau_j^c < \infty}\hat{B}_{ji} \\
  \mbox{subject to}& \\
  \hat{B}_{ji} \leq 0 \, \forall \,j& \\
  \hat{\gamma}_c \leq 0 \, \forall \,c& \\
  \log \left[ \exp{\hat{\gamma}_c} + \prod_{j;\tau_j \leq \tau_i} \left( 1 - w_j^c+ w_j^c \exp{\hat{B}_{ji}} \right) \right]& \leq 0 \, \forall \,c.
\end{align*}

\xhdr{Network sparsity}
In general, social networks are sparse in a sense that on average nodes are
connected to a constant number rather than a constant fraction of other nodes
in the network. To encourage a sparse MLE solution, an $l_1$ penalty term can
be added to the original (pre-convexification) log-likelihood function, making
the objective function
\begin{align*}
  -\log L_i(A_{:,i}|D) + \rho \sum_{j=1}^N |A_{ji}|
\end{align*}
where $\rho$ is the sparsity parameter.  Experimentation has indicated that
including this penalty function dramatically increases the performance of the
method; however, if we apply the same convexification process to this new
augmented objective function the resulting function is
\begin{align*}
  \sum_{c\in D; t_i^c < \infty}-\hat{\gamma}_c-\sum_{c\in D; t_i^c = \infty}\sum_{j \in c; t_j^c < \infty}\hat{B}_{ji} - \rho \sum_{j=1}^N\exp \hat{B}_{ji},
\end{align*}
which is concave and makes the whole problem non-convex.  Instead, we propose
the use of the penalty function $\rho\sum_{j=1}^N\frac{1}{1-A_{ji}}$.  This penalty
function still promotes a sparse solution, and even though we no longer have a
geometric program, we can convexify the objective function and so the global
convexity is preserved:
$$
  \sum_{c\in D; t_i^c < \infty}-\hat{\gamma}_c-\sum_{c\in D; t_i^c = \infty}\sum_{j \in c; t_j^c < \infty}\hat{B}_{ji} + \rho \sum_{j=1}^N\exp \left(-\hat{B}_{ji}\right).
$$

\xhdr{Implementation}  We use the SNOPT7 library to solve the likelihood
optimization. We break the network inference down into a series of subproblems
corresponding to the inference of the inbound edges of each node.
Special concern is needed for the sparsity penalty function.  The presence of
the $l_1$ penalty function makes the method extremely effective at predicting
the presence of edges in the network, but it has the effect of distorting the
estimated edge transmission probabilities.  To correct for this, the inference
problem is first solved with the $l_1$ penalty.  Of the resulting solution, the
edge transmission probabilities that have been set zero are then restricted to
remain at zero, and the problem is then relaxed with the sparsity parameter set
to $\rho=0$.
This preserves the precision and recall of the edge location prediction of the
algorithm while still generating accurate edge transmission probability
predictions. Moreover, with the implementation described above, most 1000 node
networks can be inferred inside of 10 minutes, running on a laptop.  A freely-distributable (but non-scalable) MATLAB implementation can be found at http://snap.stanford.edu/connie.

\section{Experiments}
\label{sec:experiments}
In this section, we evaluate our network inference method, which we will refer to as \textbf{ConNIe} (\textbf{Con}vex \textbf{N}etwork \textbf{I}nferenc\textbf{e}) on a range of
datasets and network topologies.  This includes both synthetically generated
networks as well as real social networks, and both simulated and real diffusion
data.  In our experiments we focus on the \SI model as it best applies
to the real data we use.

\subsection{Synthetic data}
Each of the synthetic data experiments begins with the construction of the
network. We ran our algorithm on a directed scale-free network constructed
using the preferential attachment model~\cite{barabasi99emergence}, and also on
a Erd\H{o}s-R\'{e}nyi random graph. Both networks have 512 nodes and 1024
edges. In each case, the networks were constructed as unweighted graphs, and
then each edge $(i,j$) was assigned a uniformly random transmission probability
$A_{ij}$ between 0.05 and 1.

\xhdr{Transmission time model}  In all of our experiments, we assume that the
model $w(t)$ of transmission times is known. We experimented with various
realistic models for the transmission time~\cite{bailey75mathematical}:
exponential ($w(t)=\alpha e^{- \alpha t}$), power-law ($w(t)\propto (\alpha -
1) t^{-\alpha}$) and the Weibull distribution $\left( w(t) = \frac{k}{\alpha}
\left(\frac{x}{\alpha} \right)^{k-1} e^{ -\left(\frac{x}{\alpha}\right)^k}\right)$ as
it has been argued that Weibull distribution of $\alpha= 9.5$ and  $k = 2.3$
best describes the propagation model of the SARS outbreak in Hong
Kong~\cite{wallinga04epidemic}. Notice that our model does not make any
assumption about the structure of $w(t)$. For example, our approach can handle
the exponential and power-law that both have a mode at 0 and monotonically
decrease in $t$, as well as the Weibull distribution which can have a mode at
any value.

We generate cascades by first selecting a random starting node of the
infection. From there, the infection is propagated to other nodes until no new
infections occur: an infected node $i$ transmits the infection to uninfected
$j$ with probability $A_{ij}$, and if transmission occurs then the propagation
time $t$ is sampled according to the distribution $w(t)$.
The cascade is then given to the algorithm in the form of a series of
timestamps corresponding to when each node was infected.  Not to make the
problem too easy we generate enough cascades so that $99\%$ of all edges of the
network transmitted at least one infection. The number of cascades needed for
this depends on the underlying network. Overall, we generate on the same
order of cascades as there are nodes in the network.

\xhdr{Quantifying performance}  To assess the performance of ConNIe, we
consider both the accuracy of the edge prediction, as well as the accuracy of
edge transmission probability.  For edge prediction, we recorded the precision
and recall of the algorithm. We simply vary the value of $\rho$ to obtain
networks on different numbers of edges and then for each such inferred network
we compute precision (the number of correctly inferred edges divided by the
total number of inferred edges), and recall (the number of correctly inferred
edges divided by the total number of edges in the unobserved network). For
large values of $\rho$ inferred networks have high precision but low recall,
while for low values of $\rho$ the precision will be poor but the recall will
be high.

To assess the accuracy of the estimated edge transmission probabilities
$A_{ij}$, we compute the mean-square error (MSE).  The MSE is taken over the
union of potential edge positions (node pairs) where there is an edge in the
latent network, and the edge positions in which the algorithm has predicted the
presence of an edge. For potential edge locations with no edge present, the
weight is set to 0.

\xhdr{Comparison to other methods} We compare our approach to NetInf which is
an iterative algorithm based on submodular function
optimization~\cite{gomez10netinf}. NetInf first reconstructs the most likely
structure of each cascade, and then based on this reconstruction, it selects the next most
likely edge of the social network. The algorithm assumes that the weights of
all edges have the same constant value (i.e., all nonzero $A_{ij}$ have the
same value). To apply this algorithm to the problem we are considering, we simply first use the NetInf to infer the network structure and then estimate the edge transmission probabilities $A_{ij}$ by simply counting the fraction of times it was predicted that a cascade propagated along the edge
$(i,j)$.

\begin{figure}[t]
\begin{center}
\subfigure[PR Curve (PL)]{\includegraphics[scale=.55]{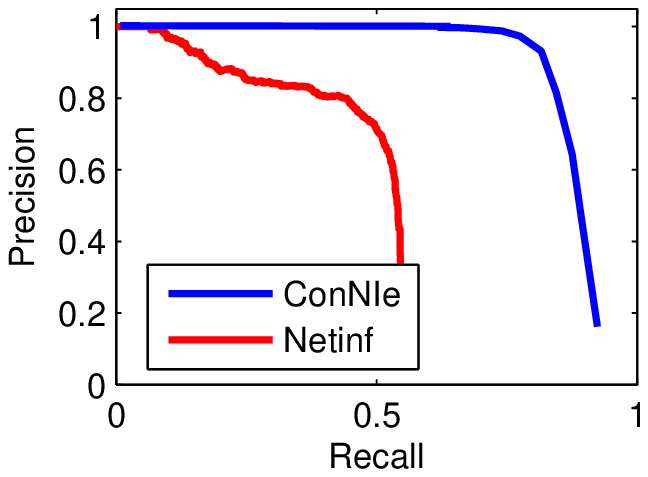}}
\subfigure[PR Curve (Exp)]{\includegraphics[scale=.55]{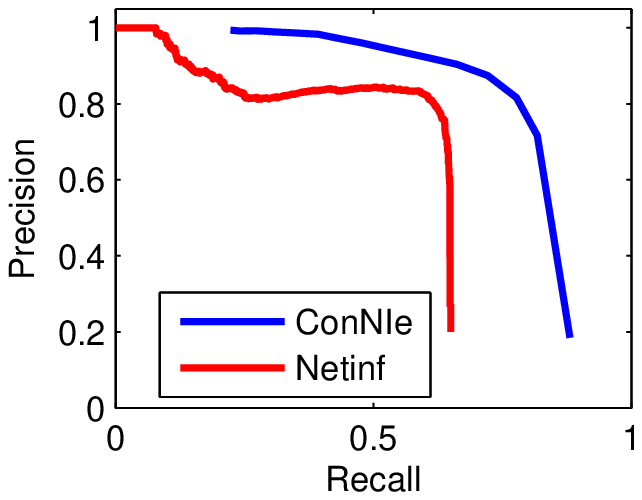}}
\subfigure[PR Curve (WB)]{\includegraphics[scale=.43]{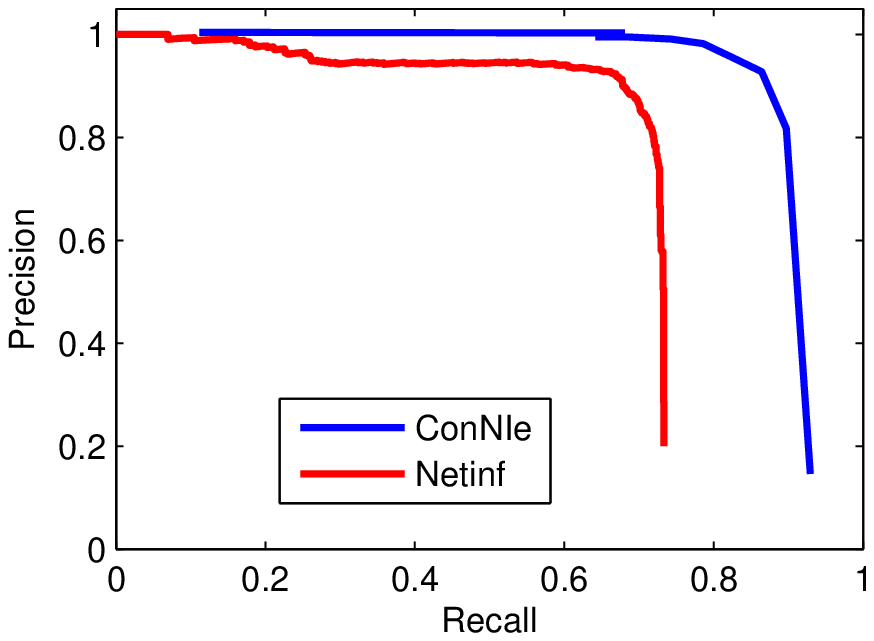}}
\subfigure[MSE (PL)]{ \includegraphics[scale=.55]{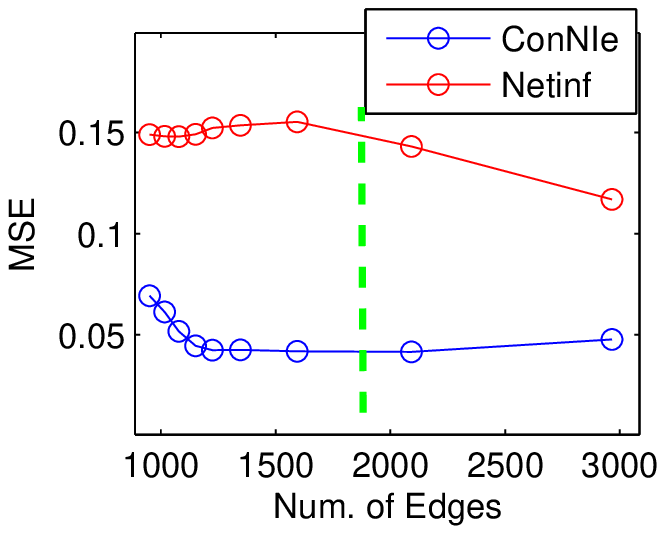}}
\subfigure[MSE (Exp)]{\includegraphics[scale=.55]{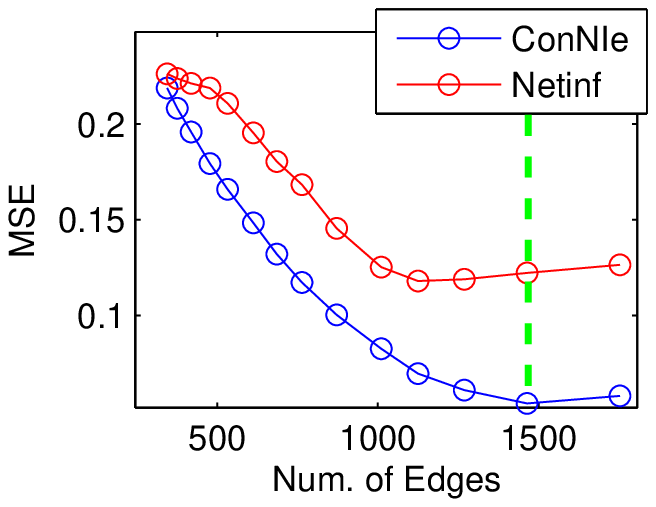}}
\subfigure[MSE (WB)]{\includegraphics[scale=.43]{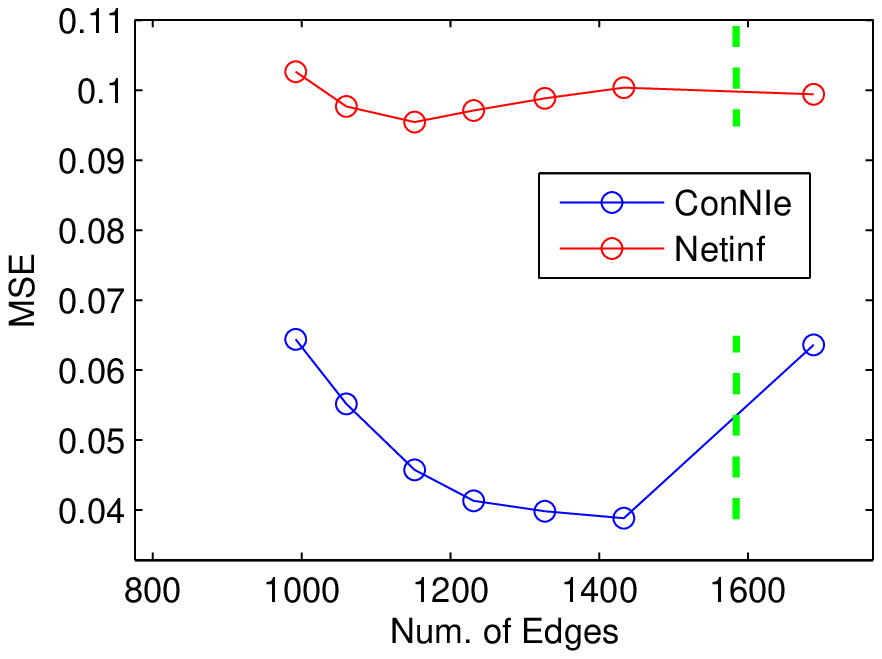}}
%\subfigure[Random graph, (PL)]{\includegraphics[scale=.23]{randPL_PR.eps}}
%\subfigure[Random graph, (Exp)]{\includegraphics[scale=.27]{randExp_PR.eps}}
%\subfigure[Random graph, (WB)]{\includegraphics[scale=.26]{randW_PR.eps}}
\vspace{-4mm}
\caption{(a)-(c): Precision and recall of ConNIe compared to NetInf for the \SI
diffusion model, run on a synthetical scale-free graph with synthetically generated cascades.  Transmission time models used are power law (PL), exponential (Exp), and Weibull (WB).
All networks contain 512 nodes, and the weight of each edge was sampled from a uniform random
distribution between 0 and 1.  For the MLE method, the PR curves were generated by varying the sparsity
parameter $\rho$ between 0 and 1000. 
(d)-(f): Mean square error of the edge transmission probability of the two algorithms.
The dotted green line indicates the number of edges in the true network. 
\label{fig:synPRCurves}}
\vspace{-5mm}
\end{center}
\end{figure}

\begin{figure}[h]
\begin{center}
\subfigure[PR Break-even (PL)]{\includegraphics[scale=.55]{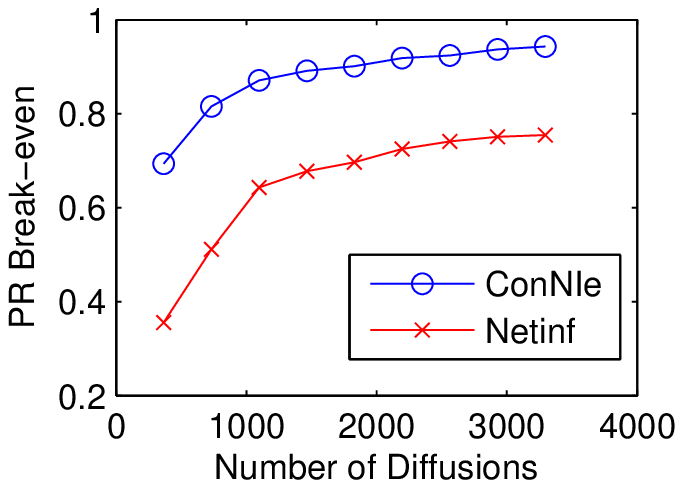}}
\subfigure[PR Break-even (EXP)]{\includegraphics[scale=.55]{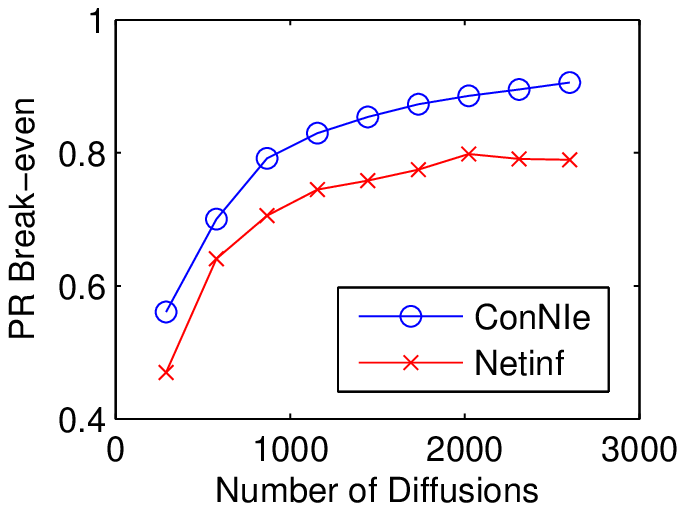}}
\subfigure[MSE (PL)]{\includegraphics[scale=.55]{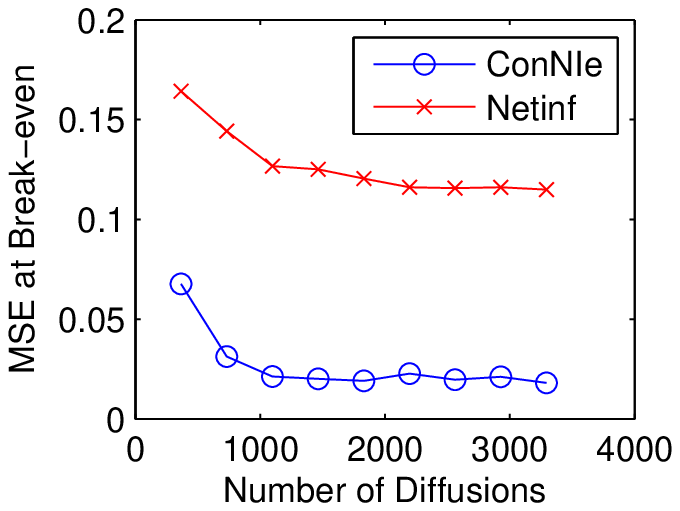}}
%\subfigure[AUC (PL) ]{\includegraphics[scale=.25]{sd_pl3.eps}}
\subfigure[MSE (EXP)]{\includegraphics[scale=.55]{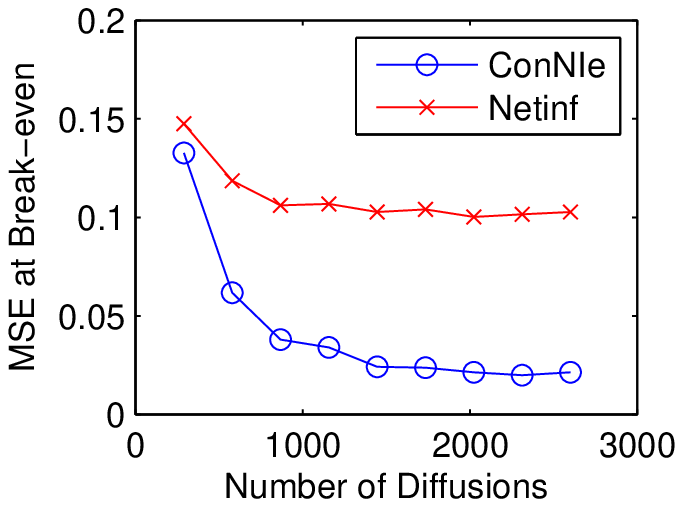}}
\subfigure[PR vs. Noise/Signal]{\includegraphics[scale=.55]{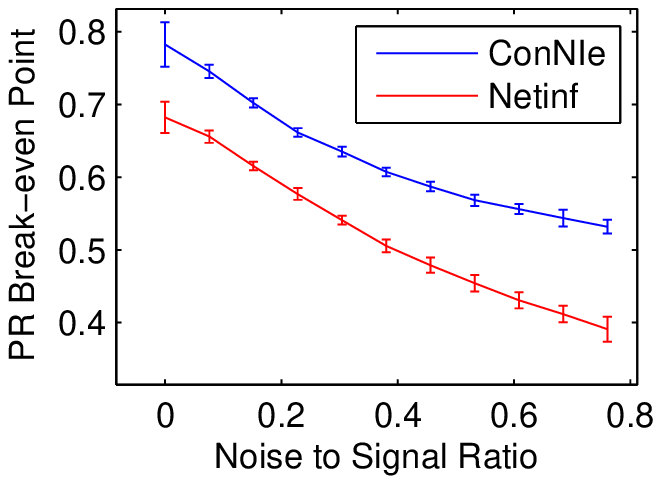}}
\subfigure[Runtime vs. Network Size]{\includegraphics[scale=.5]{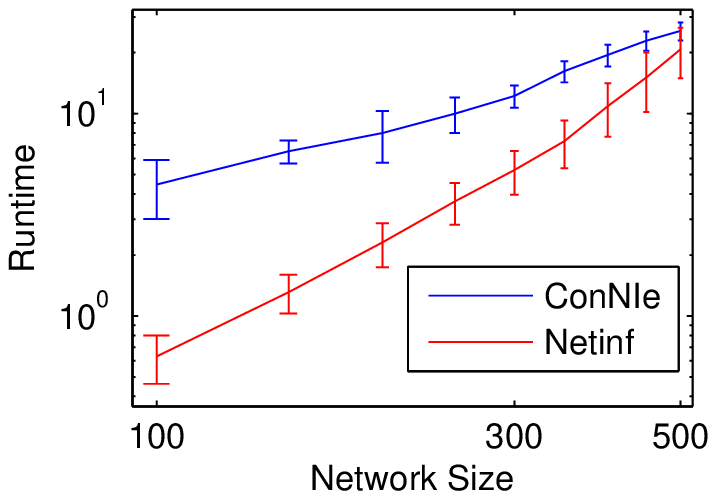}}
%\subfigure[AUC (EXP) ]{\includegraphics[scale=.25]{sd_exp3.eps}}
%\subfigure[PR Break-even point (W)]{\includegraphics[scale=.25]{sd_w1.eps}}
%\subfigure[MSE at Break-even (W)]{\includegraphics[scale=.25]{sd_w2.eps}}
%\subfigure[AUC (W) ]{\includegraphics[scale=.25]{sd_w3.eps}}
\vspace{-4mm}
\caption{(a)-(b): Precision-Recall break-even point for the two methods as a function of the number of observed cascades, with a power law (PL) and exponential (EXP) transmission distribution. (c)-(d): Mean Square Error at the PR-Break-even point as a function of the number of observed cascades.  (e)  PR Break-even point versus the perturbation size applied to the infection times.}
\label{fig:synRobust}
\vspace{-5mm}
\end{center}
\end{figure}

Figure~\ref{fig:synPRCurves} shows the precision-recall curves for the scale-free
synthetic network with the three transmission models $w(t)$. The results for the Erd\H{o}s-R\'{e}nyi random graph were omitted due to space restrictions, but they were very similar.  Notice our approach achieves the break even point (point where precision equals recall) well above
0.85. This is a notable result: we were especially careful not to
generate too many cascades, since more cascades mean more evidence that  makes
the problem easier. Also in Figure~\ref{fig:synPRCurves} we plot
the Mean Squared Error of the estimates of the edge transmission probability
$A_{ij}$ as a function of the number of edges in the inferred network. The green vertical
line indicates the point where the inferred network contains the same number of
edges as the real network. Notice that ConNIe estimates the edge weights
with error less than 0.05, which is more than a factor of two smaller than the
error of the NetInf algorithm.  This, of course, is expected as NetInf assumes the network edge weights are homogeneous, which is not the case.

We also tested the robustness of our algorithm.  Figure ~\ref{fig:synRobust} shows the accuracy (Precision-Recall break-even point as well as edge MSE) as a function of the number of observed diffusions, as well as the effect of noise in the infection times.  Noise was added to the cascades by adding independent normally distribution perturbations to each of the observed infection times, and the noise to signal ratio was calculated as the average perturbation over the average infection transmission time.  The plot shows that ConNIe is robust against such perturbations, as it can still accurately infer the network with noise to signal ratios as high as 0.4.

\subsection{Experiments on Real data}

\xhdr{Real social networks}
We also experiment with three real-world networks. First, we consider a small
collaboration network between 379 scientists doing research on networks.
Second, we experiment on a real email social network of 593 nodes and 2824
edges that is based on the email communication in a small European research
institute.

For the edges in the collaboration network we simply randomly assigned their
edge transmission probabilities. For the email network, the number of emails
sent from a person $i$ to a person $j$ indicates the connection strength.
Let there be a rumor cascading through a network, and assume the
probability that any one email contains this rumor is fixed at $\xi$. Then if
person $i$ sent person $j$  $m_{ij}$ emails, the probability of $i$ infecting
$j$ with the rumor is $A_{ij}=1-(1-\phi)(1 - \xi)^{m_{ij}}$. The parameter
$\phi$ simply enforces a minimum edge weight between the pairs who have
exchanged least one email.  We set $\xi = .001$ and $\phi=.05$.

\begin{figure}[h]
\begin{center}
\begin{tabular}{ccc}
  \begin{minipage}{0.3\textwidth}
  \begin{tabular}{cc}
	\includegraphics[scale=.7]{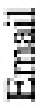} &
  	\includegraphics[scale=.45]{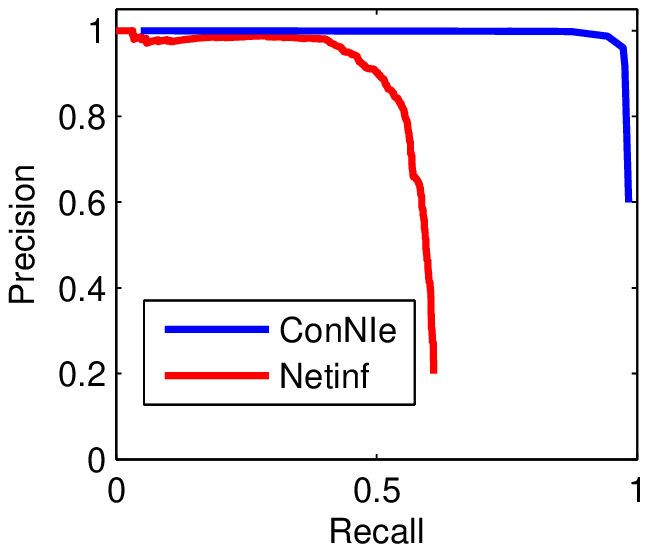} \\
	\includegraphics[scale=.7]{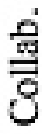} &
 	\includegraphics[scale=.45]{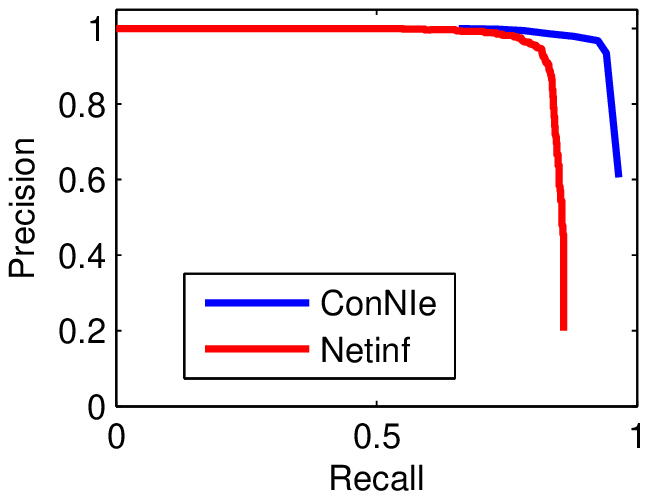}
  \end{tabular}
  \end{minipage}
  &
  \begin{minipage}{0.25\textwidth}
  \includegraphics[scale=.5]{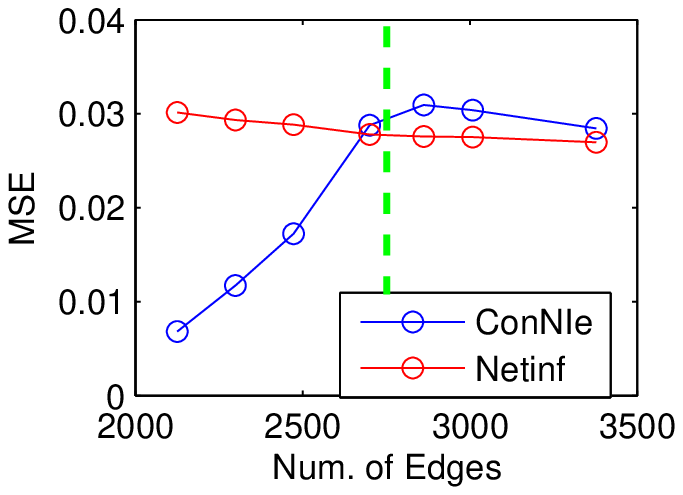}
  \includegraphics[scale=.5]{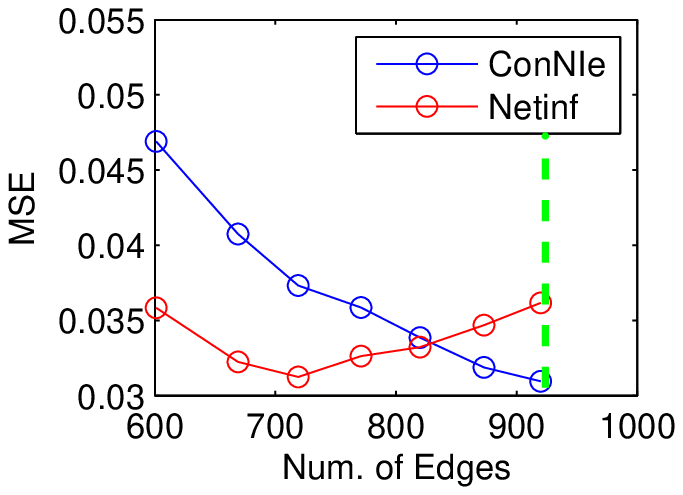}
  \end{minipage}
  &
  \begin{minipage}{0.45\textwidth}
  \includegraphics[scale=.6]{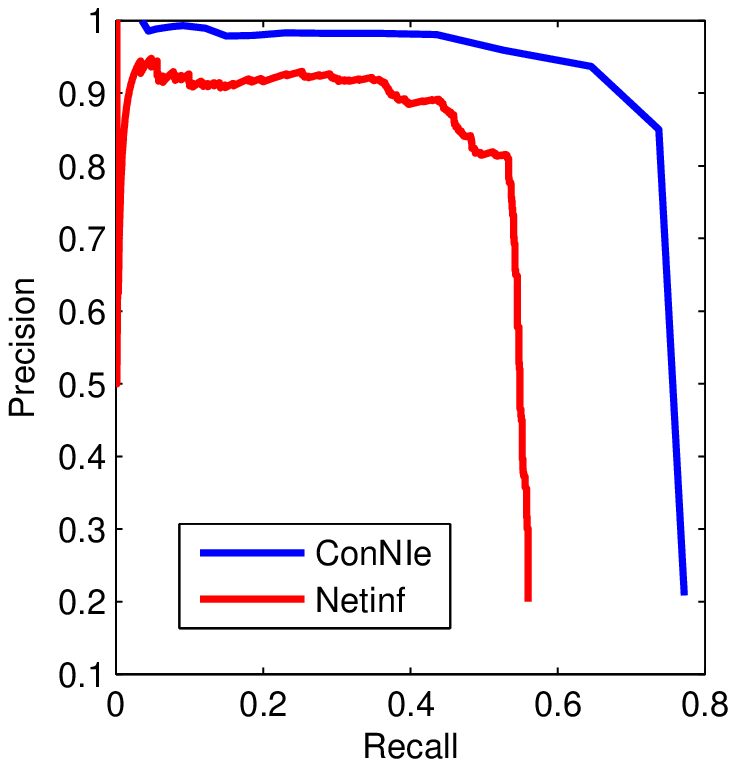}
  \end{minipage}\\
  \quad Network estimation & Edge weight error & Recommendation network\\
\end{tabular}
\caption{ The precision-recall curve of the network estimation and the mean-square error (left) of
predicted transmission probabilities as a function of number edges being
predicted (middle). Top row shows the results for the email network, and the bottom row
for the collaboration network. (Right) Precision-recall curve on inferring a real recommendation network
based on real product recommendation data. 
\label{fig:emailResults}}
\end{center}
\vspace{-5mm}
\end{figure}

For the email network we generated cascades using the power-law transmission
time model, while for the collaboration network we used the Weibull
distribution for sampling transmission times. We then ran the network inference
on cascades, and Figure~\ref{fig:emailResults} gives the results. Similarly as
with synthetic networks our approach achieves break even points of around 0.95
on both datasets. Moreover, the edge transmission probability estimation error
is less than 0.03. This is ideal: our method is
capable of near perfect recovery of the underlying social network over which a
relatively small number of contagions diffused.

\xhdr{Real social networks and real cascades} Last, we investigate a large
person-to-person recommendation network, consisting of four million people who
made sixteen million recommendations on half a million
products~\cite{jure07viral}. People generate cascades as follows: a node
(person) $v$ buys product $p$ at time $t$, and then recommends it to nodes
$\{w_1, \dots, w_n\}$. These nodes $w_i$ can then buy the product (with the
option to recommend it to others). We trace cascades of purchases on a small
subset of the data. We consider a recommendation network of 275 users and 1522
edges and a set of 5,767 recommendations on 625 different products between a
set of these users. Since the edge transmission model is unknown we model it
with a power-law distribution with parameter $\alpha=2$.

We present the results in rightmost plot of Figure~\ref{fig:emailResults}. Our
approach is able to recover the underlying social network surprisingly
accurately. The break even point of our approach is 0.74 while NetInf scores 0.55.
Moreover, we also note that our approach took less than 20 seconds to infer this
network.
Since there are no ground truth edge transmission probabilities
for us to compare against, we can not compute the error of edge weight
estimation.

%\section{Discussion}
%\label{sec:discussion}
%\input{050discussion}

\section{Conclusion}
\label{sec:conclusion}
We have presented a general solution to the problem of inferring latent social
networks from the network diffusion data.  We formulated a maximum likelihood
problem and by solving an equivalent convex problem, we can guarantee the
optimality of the solution. Furthermore, the $l_1$ regularization can be used
to enforce a sparse solution while still preserving convexity. We evaluated our
algorithm on a wide set of synthetic and real-world networks with several different
cascade propagation models.  We found our method to be more general and robust
than the competing approaches. Experiments reveal that our method
near-perfectly recovers the underlying network structure as well as the
parameters of the edge transmission model. Moreover, our approach scales well
as it can infer optimal networks on thousand nodes in a matter of minutes.

One possible venue for future work is to also include learning the parameters
of the underlying model of diffusion times $w(t)$. It would be fruitful to
apply our approach to other datasets, like the spread of a news story breaking
across the blogosphere, a SARS outbreak, or a new marketing campaign on a
social networking website, and to extend it to additional models of diffusion.
By inferring and modeling the structure of such latent social networks, we can
gain insight into positions and roles various nodes play in the diffusion
process and assess the range of influence of nodes in the network.

\xhdr{Acknowledgements}
This research was supported in part by NSF grants
CNS-1010921,  % social contagion with Madhav
IIS-1016909,    % memetracking with Jon
LLNL grant B590105, % LLNL with Tina
the Albert Yu and Mary Bechmann Foundation, IBM, Lightspeed, Microsoft and Yahoo.

\bibliography{netinf2}
\bibliographystyle{abbrv}

%\appendix
%\section{Appendix}
%\label{sec:appendix}
%\input{070appendix}

\end{document}